\documentstyle[preprint,epsfig,tighten,aps]{revtex}

\def\ut#1{\rlap{\lower1ex\hbox{$\sim$}}{#1}}

\def\SU{{\rm SU}}

\newcommand{\pic}[5]{\raisebox{#3pt}{
\hspace{#4pt}\psfig{file=#1.eps,height=#2pt,silent=}\hspace{#5pt}}}

\preprint{\vbox{\baselineskip=12pt
\rightline{CGPG-98/4-4}}}

\begin{document}
\draft
\title{Discrete Space-Time Volume for\\
3-Dimensional BF Theory and Quantum Gravity}
\author {Laurent Freidel\thanks{E-mail address: 
freidel@phys.psu.edu}${}^{1,2}$
and Kirill Krasnov\thanks{E-mail address:
krasnov@phys.psu.edu}${}^1$
}
\address{1. Center for Gravitational Physics and Geometry \\
Department of Physics, The Pennsylvania State University \\
University Park, PA 16802, USA}

\address{2. Laboratoire de Physique Th\'eorique ENSLAPP \\
Ecole Normale Sup\'erieure de Lyon \\
46, all\'ee d'Italie, 69364 Lyon Cedex 07, France}
\maketitle

\begin{abstract}
The Turaev-Viro state sum invariant is known to give the transition 
amplitude for the three dimensional BF theory with cosmological 
term, and its deformation parameter $\hbar$ is related with the 
cosmological constant via $\hbar=\sqrt{\Lambda}$. This suggests a way 
to find the expectation value of the spacetime volume by differentiating
the Turaev-Viro amplitude with respect to the cosmological constant.
Using this idea, we find an explicit expression for the spacetime
volume in BF theory. According to our results, each labelled 
triangulation carries a volume that depends on the labelling spins. 
This volume is explicitly discrete. We also show how the Turaev-Viro model 
can be used to obtain the spacetime volume for (2+1) dimensional
quantum gravity.
\end{abstract}

The loop approach to quantum gravity \cite{Review} leads to a fascinating
picture in which the structure of geometry
at the Planck scale is fundamentally different from that of 
the usual Riemannian geometry. The elementary excitations of the
quantum geometry which arises are one-dimensional, ``loop-like'', rather
than particle-like. The spectra of operators corresponding to such
geometrical quantities as length, area and volume become
quantized \cite{Quantized}. The ``size'' of the corresponding quanta
of geometry is proportional to the Planck length. 

However, so far, almost all the progress that has been made within 
this approach concerns space rather than spacetime 
aspects of quantum geometry. In this paper we propose a new set of ideas 
that can be used to study the {\it spacetime} quantum geometry. 
More precisely, using the path integral techniques,
we propose a way to define the {\it spacetime volume} in the 
quantum theory. We do this in the context of BF theory and Euclidean 
quantum gravity in three spacetime dimensions. The simplicity of these 
(classically) closely related theories allows us 
to obtain an explicit expression for 
the spacetime volume. 

There exists many ways to quantize 3-dimensional gravity (for a review
see \cite{Carlip}). In this paper we concentrate on the approach pioneered 
by Ponzano and Regge \cite{PR}. This approach is intimately related with 
the loop quantum gravity approach. Indeed, using the ideas described by Baez 
\cite{SpinFoams} (see also \cite{Rovelli}), 
one can think of the Ponzano-Regge model as
a covariant (spacetime) version of loop quantum gravity in (2+1) dimensions.
The Ponzano-Regge approach, as we explain below, is very close in spirit 
to the usual path integral approach, and, thus, is best suited for a 
study of spacetime aspects of the theory. 

Although the Ponzano-Regge model gives a quantization of BF theory,
not gravity, the two are closely related in three dimensions,
at least as classical theories. Thus, one can hope to  
use BF theory to study certain aspects of 3-d quantum gravity. We 
will first derive an expression for the spacetime volume in BF theory, and
comment on how BF theory can be used to obtain the spacetime volume
for gravity. Let us describe the Ponzano-Regge
model (or, more precisely, its generalization proposed
by Turaev and Viro, see {\it e.g.} \cite{TV}) in more details. 
Let us consider the theory whose action is given by
\begin{equation}\label{action}
S_\Lambda[A,E] = -\int_{\cal M} {\rm Tr}\left(E\wedge F + 
{\Lambda\over 12} E\wedge E\wedge E\right),
\end{equation}
where $\cal M$ is assumed to be a three-dimensional orientable
manifold. The action is a functional of an $\SU(2)$ connection $A$,
whose curvature form is denoted by $F$,
and a 1-form $E$, which takes values in the Lie algebra of
$\SU(2)$. Thus, the action (\ref{action}) is that of BF
theory in 3d, with $E$ field playing the role of $B$, and with
an additional ``cosmological term'' added to the usual BF action.
The relation to gravity in 3d is as follows.
Having the one-form $E$, one can construct from it
a real metric of Euclidean signature 
\begin{equation}\label{metric}
g_{ab} = - {1\over 2}{\rm Tr}(E_a E_b).
\end{equation}
For our conventions on indices and others see the Appendix.
Thus, the $E$ field in (\ref{action}) has the interpretation
of the triad field. One of the equations of 
motion that follows from (\ref{action}) states that $A$
is the spin connection compatible with the triad $E$. 
Taking the triad $E$ to be non-degenerate and
``right-handed'', i.e., giving
a nowhere zero positive volume form (\ref{vform}), and substituting 
into (\ref{action}) the
spin connection instead of $A$, 
one gets the Euclidean Einstein-Hilbert action
\begin{equation}
{1\over2}\int_{\cal M} d^3x \sqrt{g}\,(R-2\Lambda).
\end{equation}
We use units in which $8\pi G=1$. The coefficient in front
of (\ref{action}) is chosen to yield precisely the Einstein-Hilbert action
after the elimination of $A$. Thus, on configurations of $E$ field 
that are non-degenerate and right-handed, 
(\ref{action}) is equivalent to 
Einstein-Hilbert action, and $\Lambda$ in (\ref{action}) is 
precisely the cosmological constant. This means that all
solutions of Einstein's equations can be obtained from
the solutions of BF theory. However, BF theory has more
solutions than gravity. For example, in BF theory one can have
configurations of $E$ field that give negative or zero value of the
volume form at some points in spacetime. This does not cause any
problems classically, for one can always restrict oneself to
the sector where metric is non-degenerate and the volume 
form is positive. However, as we shall see below, this does
cause problems in the quantum theory: the amplitudes of
configurations with positive volume form interfere with the
amplitudes of negative volume configurations. Thus, as we shall see,
the quantum models corresponding to BF theory and gravity 
are rather different.

Let us first consider the quantization of BF theory.
The Turaev-Viro model gives a way to calculate
the vacuum-vacuum transition amplitude of this theory,
i.e., the path integral
\begin{equation}\label{z}
Z(\Lambda,{\cal M}) = \int {\cal D}E \,{\cal D}A \, e^{iS_\Lambda[A,E]}.
\end{equation}
Note that we consider the {\it transition amplitude} 
of BF theory, {\it not} the
partition function, which would be given by (\ref{z})
without the $i$ in the exponential. As we shall see below,
this is the transition amplitude that is related to the Turaev-Viro
state sum invariant. 
In this paper we consider only vacuum-vacuum amplitudes. Although
the Turaev-Viro model can be used to calculate more general
amplitudes between non-trivial initial and final states, we
will not use this aspect of the model here.
We consider the version of the model formulated on
a triangulated manifold. 
Thus, let us fix a triangulation $\Delta$ of $\cal M$.
Let us label the edges, for which we will employ the 
notation $e$, by irreducible representations of the
quantum group $(\SU(2))_q$, where $q$ is a root of unity
\begin{equation}\label{q}
q = e^{{2\pi i\over k}} \equiv e^{i\hbar}.
\end{equation} 
Later we will relate the parameter $\hbar$ with 
the cosmological constant $\Lambda$. The irreducible representations
of $(\SU(2))_q$ are labelled by half-integers (spins) $j$ satisfying
$j \leq (k-2)/2$. Thus, we associate a spin $j_e$ to each 
edge $e$. The vacuum-vacuum transition amplitude of the theory is then
given by the following expression (see, {\it e.g.} \cite{TV}):
\begin{equation}\label{tv}
{\rm TV}(q,\Delta) = \kappa^V\,\sum_{j_e} \prod_e {\rm dim}_q(j_e) 
\prod_t (6j)_q,
\end{equation}
where $\kappa$ and ${\rm dim}_q(j)$ are defined by 
(\ref{kappa}),(\ref{qdim}) correspondingly, and 
$V$ is the number of vertices in $\Delta$.
The last product in (\ref{tv}) is taken over tetrahedra $t$ of $\Delta$, 
and $(6j)_q$ is the normalized quantum (6j)-symbol constructed from the 6 spins
labelling the edges of $t$. It turns out that (\ref{tv}) is
independent of the triangulation $\Delta$ and gives a topological 
invariant of $\cal M$: ${\rm TV}(q,\Delta)={\rm TV}(q,\cal M)$.

The construction that interprets the Turaev-Viro invariant (\ref{tv}) 
as the vacuum-vacuum transition amplitude of the theory defined
by (\ref{action}) is as follows. It has been proved (see {\it e.g.}
\cite{Roberts} and the works cited therein) that (\ref{tv}) is equal
to the squared absolute value of the Chern-Simons amplitude
\begin{equation}\label{tv-cs}
{\rm TV}(q,{\cal M}) = |{\rm CS}(k,{\cal M})|^2,
\end{equation}
with the level of Chern-Simons theory being equal to $k$
from (\ref{q}). It is known, however, that the action
(\ref{action}) can be written as a difference of two copies of
Chern-Simons action
\begin{equation}
S_{\rm CS}(A) = {k\over 4\pi} \int_{\cal M}
{\rm Tr}\left( A\wedge dA + {2\over 3} A\wedge A\wedge A\right).
\end{equation} 
Indeed, note that
\begin{equation}\label{two}
S(A+\lambda E) - S(A-\lambda E) = {k\lambda\over\pi} \int_{\cal M} 
{\rm Tr}\left( E\wedge F + {\lambda^2\over 3} E\wedge E\wedge E\right),
\end{equation}
where $\lambda$ is a real parameter. Thus, (\ref{two}) is
equal to (\ref{action}) if 
\begin{equation}\label{rel}
\lambda = - \sqrt{\Lambda}/2, \qquad k = {2\pi\over\sqrt{\Lambda}},
\qquad {\rm or} \qquad \hbar = \sqrt{\Lambda}.
\end{equation}
This relates the deformation parameter $q$ of the Turaev-Viro
model to the cosmological constant $\Lambda$, in the
case of positive $\Lambda$, and proves
that the Turaev-Viro amplitude is proportional to the 
vacuum-vacuum transition amplitude of the theory defined by
(\ref{action}):
\begin{equation}
{\rm TV}(q,{\cal M}) \sim Z(\Lambda,{\cal M}).
\end{equation}

Let us now discuss the relation between quantum BF theory and  
gravity. We note that, although classically BF theory and gravity
are equivalent (they lead to the same equations of motion), the 
transition amplitude of BF theory 
(\ref{z}) is {\it not} the transition amplitude for gravity. Indeed, 
in the case of gravity one has to perform the path integral of
the exponentiated action only over non-degenerate metric configurations
that define a positive spacetime volume:
\begin{equation}
Z_{gr}(\Lambda,{\cal M}) = \int_{{\rm Vol}(E) > 0} 
{\cal D}E \,{\cal D}A \, e^{iS_\Lambda[A,E]}.
\end{equation}
However, the path integral
in (\ref{z}) is taken over {\it all} configurations of $E$ field,
even those that are degenerate or define a negative volume. 
To see this, let us note
that if the field $E$ defines an everywhere positive volume, then $-E$ gives a
negative volume, and both are summed over in the path integral
(\ref{z}). Thus, the path integral includes contributions from
both configurations of $E$ field that give an everywhere positive
and the ones that given an everywhere negative volume. 
Moreover, the path integral (\ref{z}) takes into
account also degenerate configurations of $E$ in which the volume form is
negative at some points in spacetime and positive at other. As we further 
discuss below, these are these degenerate configurations that may cause
the quantum BF theory to be drastically different from gravity, if
it turns out that they dominate the path integral.
To further illustrate the difference between the quantum BF theory and
gravity, let us write a heuristic expressions for the path integrals
corresponding to the two theories. For the transition amplitude in 
gravity one can formally write:
\begin{equation}
\int {\cal D}g \prod_{x\in\cal M} e^{i{\cal L}_{gr}(x)},
\end{equation}
where the path integral is taken over non-degenerate configurations
of metric $g$ and ${\cal L}_{gr}(x)$ is the Lagrangian of gravity.
In the path integral of the quantum BF theory, one takes into account
all, even highly degenerate configurations of the $E$ field. Thus,
heuristically, its transition amplitude is given by
\begin{equation}
\int_{{\rm Vol}(E) \geq 0} 
{\cal D}E {\cal D}A \prod_{x\in\cal M} \cos({\cal L}_{BF}(x)) =
\int {\cal D}g \prod_{x\in\cal M} \cos(i{\cal L}_{gr}(x)).
\end{equation}
Here the integral is taken over configurations of $E$ field that
give a non-negative volume, and the presence of $\cos$ 
accounts for the fact that one sums at each point both 
over positive and negative volumes. The presence of $\cos$ here
is also reminiscent of the fact (see \cite{PR}) that the Ponzano-Regge 
amplitude, which is given by the product of (6j)-symbols
(see (\ref{amplitude}) below), in the limit of large 
spins $j_e$ has the asymptotics of the cosine of the
Regge calculus version of the Einstein-Hilbert action. 
Thus, to summarize,
the quantum BF theory may well be drastically different from gravity
because it takes into account highly degenerate, classically
forbidden field configurations. However, the two theories
would be related if there exists a phase of the quantum BF
theory in which the path integral is dominated by non-degenerate
configurations of $E$ field, that is, configurations that
have everywhere positive or negative volume. In this phase
one would effectively have to consider only everywhere 
positive or everywhere negative volume configurations.
In the transformation $E\to -E$ the BF action
(\ref{action}) changes its sign. 
Thus, in this phase
\begin{equation}\label{rel1}
Z(\Lambda,{\cal M}) = Z_{gr}(\Lambda,{\cal M}) +
\overline{Z_{gr}(\Lambda,{\cal M})},
\end{equation}
where overline denotes complex conjugation.
In this phase there exists a relation between the 
spacetime volume in BF theory and gravity. We shall
discuss this after we obtain an expression for the
volume in the quantum BF theory.

Let us now study the spacetime volume in 
BF theory. Note that the expectation value of 
the volume is given simply by the derivative of
the amplitude (\ref{z}) with respect to $(-i\Lambda)$
\begin{eqnarray}\label{der}
\langle{\rm Vol}\rangle = 
{\int {\cal D}E {\cal D}A \, {\rm Vol}({\cal M}) e^{iS}
\over \int {\cal D}E {\cal D}A \, e^{iS}} = 
i\,{\partial \ln{Z(\Lambda)}\over\partial \Lambda}, \\
{\rm Vol}({\cal M}) = \int_{\cal M}
{1\over 12} \tilde{\varepsilon}^{abc}
{\rm Tr}(E_a E_b E_c).
\nonumber
\end{eqnarray}
Thus, the expectation value of the volume of $\cal M$
in BF theory can be obtained by differentiating the Turaev-Viro
amplitude with respect to $\Lambda$. Here we
calculate this derivative and find an explicit 
expression for the spacetime volume. For
simplicity, we will study only the
volume in the quantum theory with zero cosmological
constant. This can be obtained by first differentiating
(\ref{tv}) with respect to $\Lambda$, and then evaluating the 
result at $\Lambda=0$. Thus, interestingly, the deformation
parameter of the Turaev-Viro model $q$ 
(related with $\Lambda$ via (\ref{rel})) serves as the 
quantity conjugate to the spacetime volume.

Since the Turaev-Viro amplitude is explicitly real, and,
to find the volume, we differentiate its logarithm with
respect to $i\Lambda$, the expectation value of the volume
is {\it purely imaginary}. This is, at the first sight, surprising,
but can be easily understood by taking into account the fact
that both positive and negative volume configurations
are summed over in (\ref{z}), and that the action (\ref{action})
changes its sign when $E\to -E$.

Let us now find this volume. 
An important subtlety arises, however, if one follows the
procedure (\ref{der}). To find the derivative (\ref{der}) at $\Lambda=0$
we have to find the first order term in $\Lambda$ in the
expansion of (\ref{tv}). It is not hard to show, however,
that (\ref{tv}) has the following asymptotic expansion
in $\hbar$
\begin{equation}\label{dec}
\left({\hbar^3\over 4\pi}\right)^V\,{\rm PR}(\Delta)\,
\left(1-\hbar^2 i\langle{\rm Vol}\rangle\right),
\end{equation}
where $\rm PR$ is the amplitude of the Ponzano-Regge model 
\begin{equation}\label{pr}
{\rm PR}(\Delta) = \sum_{j_e} \prod_e {\rm dim}(j_e) \prod_t (6j),
\end{equation}
$V$ is the number of vertices in $\Delta$,
and $i\langle{\rm Vol}\rangle$ is a real quantity independent of $\hbar$.
Thus, apparently there is no term proportional to $\hbar^2$
in this expansion. The resolution of this is that the
deformation level $k$ of the Turaev-Viro model plays 
two distinct roles in the theory. First, it serves as
a regulator for the Ponzano-Regge model. Indeed, the
amplitude (\ref{pr}) diverges. It is only the combination
$(\hbar^{3V}{\rm PR})$, defined as the limit of $(\rm TV)$
as $\hbar\to 0$, that is finite. Second, the parameter $k$
(and the related to it $q$) also serves as a deformation 
parameter of the Ponzano-Regge model.
The terms of the order $O(\hbar^2)$ in (\ref{dec}) are
the ones that appear as the result of this deformation. To derive an expression
for the spacetime volume we must be concerned only with these
terms, for they describe the ``deformation'' of the vacuum-vacuum
transition amplitude occuring due to the introduction of the
cosmological constant term into the action functional. Thus,
the expectation value of the spacetime volume is what is
denoted by $\langle{\rm Vol}\rangle$ in (\ref{dec}).

There is another, equivalent way to justify the absence of
terms proportional to $\hbar^2$ in the decomposition (\ref{dec}).
As we have mentioned above, the Turaev-Viro amplitude (\ref{tv}) 
is proportional to the transition amplitude (\ref{z}).
However, the proportionality coefficient {\it depends}
on $\hbar$. Indeed, the integration over $(A+\lambda E), (A-\lambda E)$,
which is carried out to obtain $|{\rm CS}(k,{\cal M})|^2$ in (\ref{tv-cs})
and thus the Turaev-Viro amplitude, is different from the integration over $A,E$
one has to perform to obtain (\ref{z}). The difference in the 
integration measures is a power of $\hbar$. Thus, the 
amplitude (\ref{z}) and the squared absolute value of the 
amplitude of the Chern-Simons theory are proportional to
each other with the coefficient of proportionality being
a power of $\hbar$. In the discretized version
of the theory, given by the Turaev-Viro model, this
power of $\hbar$ is replaced by $\hbar^{3V}$.
Thus, this is the (divergent) Ponzano-Regge amplitude
(\ref{pr}) that gives the amplitude for BF theory without
the cosmological constant. Turaev-Viro amplitude, in the
limit of small cosmological constant, differs from the
BF amplitude by a power of $\hbar$.

These remarks being made, it is straightforward to write
down an expression for the expectation value of the volume:
\begin{eqnarray}\nonumber
i\langle{\rm Vol}\rangle_\Delta = - {\partial\over\partial\Lambda} 
\left( {{\rm TV}(\Lambda,\Delta)\over {\rm PR}(\Delta)(\hbar^3/4\pi)^V} 
\right)_{\Lambda=0} = \\
{1\over {\rm PR}(\Delta)} \sum_{j_e} i{\rm Vol}(\Delta,{\bf j}) 
\left( \prod_e {\rm dim}(j_e) \prod_t (6j)\right),
\label{vol-exp}
\end{eqnarray}
where the function ${\rm Vol}(\Delta,{\bf j})$ of the triangulation $\Delta$
and the labels ${\bf j}=\{j_e\}$ is given by
\begin{eqnarray}
i{\rm Vol}(\Delta,{\bf j}) = \nonumber
\sum_v \left( - {\partial\over\partial\Lambda}
\left( {\kappa \over (\hbar^3/4\pi)} 
\right)\right)_{\Lambda=0} +
\sum_e \left(- {\partial\ln({\rm dim}_q(j_e))\over\partial\Lambda}
\right)_{\Lambda=0}  + \\
\sum_t \left(- {\partial\ln((6j)_q)\over\partial\Lambda}
\right)_{\Lambda=0}. \label{vol-1}
\end{eqnarray}
Here $v$ stands for vertices of $\Delta$, $e$ stands for edges and
$t$ stands for tetrahedra.
We intentionally wrote the expectation value of the volume
in the form (\ref{vol-exp}) to introduce the volume ${\rm Vol}(\Delta,{\bf j})$
of a labelled triangulation, which is our main object
of interest. Indeed, (\ref{vol-exp}) has the form 
\begin{equation}
{\sum_{j_e} i{\rm Vol}(\Delta,{\bf j}) {\rm Amplitude}(\Delta,{\bf j})
\over \sum_{j_e}{\rm Amplitude}(\Delta,{\bf j})},
\end{equation}
where
\begin{equation}\label{amplitude}
{\rm Amplitude}(\Delta,{\bf j})=\prod_e {\rm dim}(j_e) 
\prod_t (6j)
\end{equation}
is the amplitude of Ponzano-Regge model.
This shows that ${\rm Vol}(\Delta,{\bf j})$
indeed has the interpretation of the volume of a labelled
triangulation.

The volume (\ref{vol-1}) has three types of contributions: (i)
from vertices; (ii) from edges; (iii) from tetrahedra. It is not
hard to calculate the first two types of them. One finds
that each vertex contributes exactly $1/12$, and each edge contributes
$j_e(j_e+1)/6$, where $j_e$ is the spin that labels the edge $e$. 
It is much more complicated to find the
tetrahedron contribution to the volume, that is, the 
derivative of $\ln((6j)_q)$ with respect to $\Lambda$.
Here we simply give the result of this calculation. The
details will appear elsewhere \cite{Laurent}. The simplest
way to describe the result is graphical. First, let us,
for each tetrahedron $t$, introduce a special graph $\Gamma$
living on the boundary of $t$. The boundary of $t$ is a triangulated
2-manifold with the topology of a sphere, and we define the 
graph $\Gamma$ to be 
the one dual to that triangulation. The graph $\Gamma$ has four
vertices and six edges, and, as a piecewise linear complex,
it is a tetrahedron. Let us labell the edges of this graph, which
are dual to the edges of the original graph, by the
same spins as those labelling the edges of $t$. Let us then
construct the spin network corresponding to the labelled $\Gamma$
(for more details on spin networks, see {\it e.g.} \cite{BaezSpinNets}).
Recall that a spin network corresponding to a labelled graph 
is a function on a certain number of copies of the gauge group --
$G^E$, where $E$ is the number of edges in the graph --
which is constructed by taking the matrix elements of the group elements 
in the representations labelling the corresponding edges, and 
contracting these matrix elements with each other at vertices
using intertwiners. In our case of $\Gamma$ being a tetrahedron,
all vertices are trivalent, that is, there are exactly three
edges meeting at each vertex. In this case intertwiners are
unique (up to an overall multiplicative constant), and given
by the usual Clebsch-Gordan coefficients. We choose the
normalized intertwiners; then the evaluation of the 
spin network on all group elements equal to the unity in 
$\SU(2)$ gives just the normalized classical $(6j)$-symbol, as
the one in (\ref{pr}). This can be represented
graphically by: 
\begin{equation}
\left(\pic{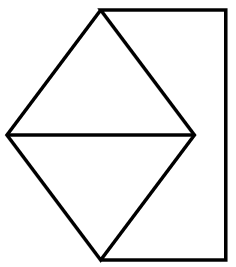}{30}{-15}{-2}{2} \right)_0 = (6j).
\end{equation}
Here the picture represents the spin network constructed
above, and $(\cdot)_0$ represents its evaluation on
the unity group elements. One gets a number that depends
on the six spins labelling the edges of the spin network.
This number is exactly the normalized classical $(6j)$-symbol.

Let us now introduce an operation that can be called {\it grasping}.
In this paper we introduce only certain basic graspings; for
more details see {\it e.g.} \cite{BN}.
Grasping will always be represented by a dashed line with open 
ends. The basic grasping is given by a line with two 
open ends. Each open end of this dashed line can be thought
of as representing the Pauli matrices $\sigma^i, i=1,2,3$ and 
the line connecting the ends represents the contraction 
$\sigma^i\otimes\sigma^i$. In this paper we use a normalization
such that each open end stands 
for $\sigma^i/\sqrt{2}$, and each dashed line stands for 
$\delta^{ij}$. The basic operation
of grasping is that each open end can grasp any of the edges 
of a spin network state. The result of this grasping is the
``insertion'' of $\sigma^i/\sqrt{2}$ into the edge of the
spin network, in the same representation as the one
labelling the edge that is being grasped. Thus, for example, 
the basic dashed line with two open ends can grasp with both
its ends one and the same edge of a certain spin network. Let
$j$ denote the spin labelling this edge. Then the grasping simply
multiplies the spin network by the value of the Casimir in the
representation $j$:
\begin{equation}
\pic{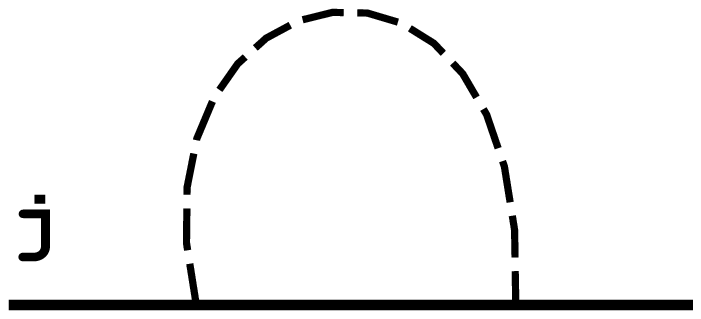}{30}{-15}{2}{2} = 2j(j+1) \pic{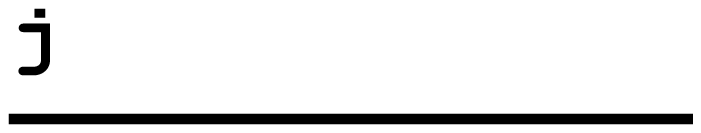}{13}{-6}{2}{2}
\end{equation}
One can also construct a more complicated grasping with three
open ends by commuting the two basic 2-open-end graspings
\begin{equation}
\pic{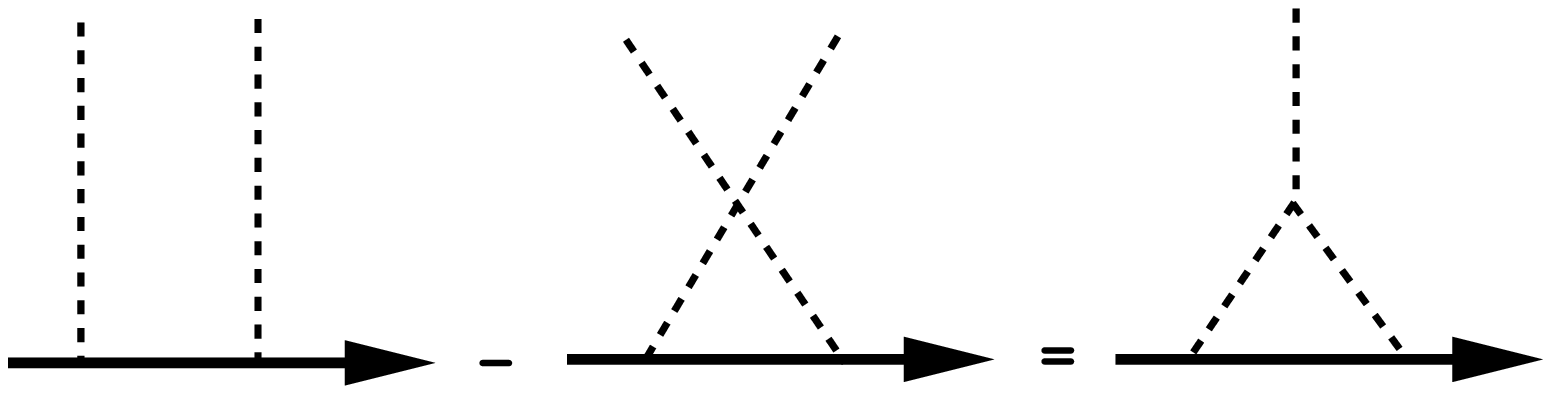}{60}{-30}{2}{2}
\end{equation}
Thus, with our normalizations, the vertex of this triple 
grasping stands for $i\sqrt{2}\varepsilon^{ijk}$. This triple
grasping is one of our main objects in what follows.

We are now in the position to describe the result of the derivative
of $(6j)_q$ symbol with respect to $\Lambda$. The result can be
represented as a sum over graspings of the edges of the spin
network constructed above:
\begin{eqnarray}
\label{tet1}
\left(- {\partial (6j)_q\over\partial\Lambda}\right)_{\Lambda=0} = \\
{1\over 16} \left( 
{1\over 24} \sum_{e,e',e''} \left\langle \pic{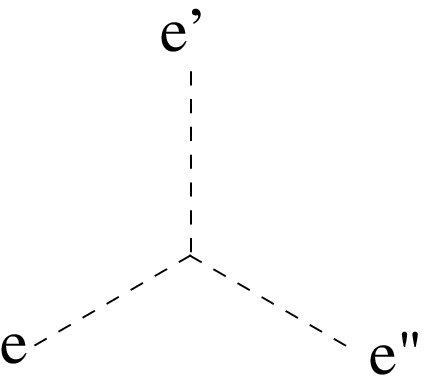}{60}{-30}{2}{2} | 
\pic{tet}{60}{-30}{2}{2} \right\rangle - {1\over 4} \sum_e \left\langle
\pic{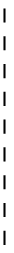}{60}{-30}{2}{2} | \pic{tet}{60}{-30}{2}{2} \right\rangle
\right)_0. \nonumber
\end{eqnarray}
Here the first sum is taken over all the graspings of the 
triples of distinct edges of the tetrahedron $t$
(there are 20 different graspings of this type, but
only 16 of them are non-zero), 
and the second sum is over all graspings of the edges of $t$
(there are 6 graspings of this type -- equal to the number of 
edges of $t$).
The result of this graspings is then evaluated on the unity
group elements to get a number. The derivative (\ref{tet1})
is a real number depending only on the spins labelling the
edges of $t$.

Thus, the final result for the spacetime volume in BF theory is 
\begin{eqnarray}\nonumber
i{\rm Vol}(\Delta,{\bf j}) = \sum_v {1\over 12} +
\sum_e {j_e(j_e+1)\over 6} + \\ \label{vol-2}
\sum_t {1\over 16} {1\over \pic{tet}{10}{-1}{-1}{0}} \left( 
{1\over 24} \sum_{e,e',e"} \left\langle \pic{grasp}{60}{-30}{2}{2} | 
\pic{tet}{60}{-30}{2}{2} \right\rangle - {1\over 4} \sum_e \left\langle
\pic{grasp2}{60}{-30}{2}{2} | \pic{tet}{60}{-30}{2}{2} \right\rangle
\right)_0.
\end{eqnarray}
Here $\pic{tet}{10}{-1}{-1}{0}$ stands for the classical $(6j)$-symbol.

It is interesting to note that not only tetrahedra $t$ of $\Delta$
contribute to the volume, but also the edges $e$ and the vertices $v$.
The contribution from the vertices is somewhat trivial -- it is constant
for each vertex. Nevertheless, when thinking about the 
triangulated manifold $\cal M$ one is forced to assign the
spacetime volume to every vertex. The contribution from edges depends on 
the spins labelling the edges. Again, this implies that each
edge of the triangulation $\Delta$ carries an intrinsic volume
that depends on its spin. The contribution from tetrahedra is more
complicated. It is given by a function that depends on the spins
labelling the edges of each tetrahedron. It is interesting that 
this picture of the spacetime volume being split into
contributions from vertices, edges and tetrahedra can be understood 
in terms of Heegard splitting of $\cal M$. Recall, that Heegard
splitting of a three-dimensional manifold $\cal M$ decomposes
$\cal M$ into three dimensional manifolds with boundaries. Then
the original manifold can be obtained by gluing these manifolds
along the boundaries. For the case of a triangulated manifold 
$\cal M$, as we have now, the Heegard splitting 
proceeds as follows. First, one constructs balls centered
at the vertices of $\Delta$. Then one connects these balls
with cylinders, whose axes of cylindrical symmetry coincide
with the edges of $\Delta$. Removing from $\cal M$ the obtained
balls and cylinders, one obtains a three-dimensional manifold with
a complicated boundary. One has to further cut this manifold
along the faces of $\Delta$. One obtains three types of ``building blocks''
that are needed to reconstruct the original manifold:
(i) balls; (ii) cylinders; (iii) spheres with four discs
removed. Each of this manifolds carries a part of the
original volume of $\cal M$. Our result (\ref{vol-2})
provides one with exactly the same picture: the volume
of $\cal M$ is concentrated in vertices (balls of the
Heegard splitting), edges (cylinders), and tetrahedra
(4-holed spheres).

The volume of each labelled triangulation is explicitly discrete,
and is given by the function (\ref{vol-2}) of the
labelling spins. This gives an important insight as to the nature of 
quantum spacetime geometry. Indeed, one of the results of 
the canonical approach to quantum gravity in three spacetime dimensions 
is that lengths of curves and areas of regions are quantized.
Our results also imply that the spacetime volume is quantized.

An important drawback of the present approach is that the spacetime
volume we have obtained is that of BF theory, not gravity. In
particular, this is the reason why the spacetime volume turns out
to be purely imaginary. Let us now discuss a relation of the 
obtained spacetime volume of BF theory with the volume in 
gravity. As we have discussed above, the quantum BF
theory may develop a phase in which non-degenerate
configurations of $E$ field dominate the path integral.
In this phase it is possible to obtain some information
about the spacetime volume in gravity from the
quantum BF theory. As we described above, in such
a phase the BF theory transition amplitude is related
to that in gravity according to (\ref{rel1}). One can
obtain a similar relation between the expectation
value of the volume in the two theories. Thus, if BF theory
is in the phase in which non-degenerate solutions dominate,
one has:
\begin{equation}
\langle{\rm Vol}\rangle_{gr} = {{\rm Re} Z_{gr}(\Lambda,{\cal M})\over
i {\rm Im} Z_{gr}(\Lambda,{\cal M})} \langle{\rm Vol}\rangle.
\end{equation}
To get this relation we have assumed that the expectation
value $\langle{\rm Vol}\rangle_{gr}$ of the volume in gravity
is real. Thus, even in this phase, to relate the volume 
in gravity with the one in BF theory
one has to know the imaginary part of the gravity amplitude.
Unfortunately, it is not possible to extract this information
from the Turaev-Viro model, which knows only about the real part
of $Z_{gr}$. Thus, one cannot extract the expectation value of 
spacetime volume of 3d gravity from BF theory. However,
it turns out to be possible to extract the expectation value
of the volume squared. Indeed, using the same arguments as
in (\ref{rel1}), and assuming that the BF theory is in the
non-degenerate phase, one can write
\begin{equation}
\langle{\rm Vol}^2\rangle Z(\Lambda,{\cal M})=
\langle{\rm Vol}^2\rangle_{gr} Z_{gr}(\Lambda,{\cal M})+
\overline{\langle{\rm Vol}^2\rangle_{gr} Z_{gr}(\Lambda,{\cal M})}.
\end{equation}
Thus, assuming that $\langle{\rm Vol}^2\rangle_{gr}$ is a 
real quantity, we obtain
\begin{equation}
\langle{\rm Vol}^2\rangle_{gr} = \langle{\rm Vol}^2\rangle.
\end{equation}
In other words, when non-degenerate fields $E$ dominate 
the path integral (\ref{z}), the expectation
value of the volume squared in Turaev-Viro model is the same as
in gravity. Thus, it can be obtained by the methods of this
paper by differentiating the Turaev-Viro amplitude with 
respect to $\Lambda$:
\begin{equation}
\langle{\rm Vol}^2\rangle_\Delta = - {\partial^2\over\partial\Lambda^2} 
\left( {{\rm TV}(\Delta)\over {\rm PR}(\Delta)(\hbar^3/4\pi)^V} 
\right)_{\Lambda=0}.
\end{equation}
This also allows one to obtain an expression for the
squared volume of a labelled triangulation. The
results of this paper indicate that this squared volume
will also be discrete, and depend only on the
spins labelling the triangulation. Thus, if there
indeed exist a phase of the quantum BF theory in which 
non-degenerate $E$ fields dominate, then one can gain
some control over the spacetime volume in (2+1) quantum
gravity using the results from BF theory. If, on the
other hand, the path integral of BF theory is always
dominated by highly degenerate solutions, as may well
be the case, then the two theories have very little to
do with each other, and the spacetime volume of
BF theory is {\it not} the same as the volume in 
gravity. At the present stage of our understanding
of quantum gravity it is hard to tell which of
this two possibilities is realized.

Let us conclude by pointing out a possible application of
the results of this paper. One of the main conceptual 
problems of quantum gravity is the absence of time in
the corresponding quantum theory. In fact, this comes about 
not just for gravity, but for any generally covariant theory,
as is, for example, BF theory discussed above.
In the canonical theory this manifests 
itself in the fact that the Hamiltonian becomes 
a constraint; in the path integral approach a
manifestation of this is that the path integral
gives not the transition amplitude between states
at different times, but the matrix element of the
projection operator on the subspace of solutions
of the Hamiltonian constraint. In the path integral
approach this arises because one is summing over
all spacetime geometries that can be put between the 
initial and final hypersurfaces. This is the basic reason
why one loses track of the time: one is summing amplitudes
of all spacetime geometries, even those which give a different
proper time separation between the initial and final hypersurfaces.
However, a possible alternative to this arises if one
constructs the path integral as a sum over labelled
triangulations, as we did above for the case of 
BF theory in three dimensions. In this case, as we found,
each labelled triangulation can be assigned the spacetime
volume carried by it. Thus, instead of summing over all
triangulations, one can consider only the sum over
triangulations having a fixed spacetime volume 
contained between the initial and final hypersurfaces.
The sum over amplitudes of such triangulations would
depend on the fixed spacetime volume. One can interprete
this spacetime volume as a measure of the time elapsed between the
initial and final hypersurfaces. Thus, an expression
for spacetime volume, as, for example, the one obtained in this
paper, allows one to introduce a natural time in the quantum theory.

{\bf Acknowledgements}: We are grateful to John Baez
for discussions and for careful reading of the early versions of this
paper. L. F. was supported by CNRS (France) and by a NATO grant.
K. K. was supported in part by the NSF
grant PHY95-14240, by Braddock fellowship from Penn State and by the 
Eberly research funds of Penn State. 

\appendix
\section{}

All the traces we use in this paper are traces in the fundamental
representation. Lower case latin letters stand for spacetime
indices: $a,b,...=1,2,3$.
The $E$ field we use is assumed to be anti-hermitian. This
explains the minus sign in (\ref{metric}). Note the
factor of $1/2$ in (\ref{metric}). This is not the 
standard choice in the gravity literature, but turns
out to be very convenient in 3d, for it allows one
to get rid of the ugly factors of $\sqrt{2}$ in some formulas.
In case $E$ has an interpretation of the triad field, the volume
form is given by 
\begin{equation}\label{vform}
{1\over 12} \tilde{\varepsilon}^{abc}
{\rm Tr}(E_a E_b E_c).
\end{equation}
Note that the volume form defined by $E$ can be both positive and negative,
and only a configuration of $E$ giving the positive volume
is a triad field of gravity.

The quantity $\kappa$ in (\ref{tv}) that, in the limit of
$\hbar\to 0$, serves as a regulator of the Ponzano-Regge model,
is defined by:
\begin{equation}\label{kappa}
\kappa = - {(q^{1/2}-q^{-1/2})^2\over 2k}. 
\end{equation}

The quantum dimension ${\rm dim}_q(j)=[2j+1]_q$, where $[n]_q$ is
the quantum number
\begin{equation}\label{qdim}
[n]_q = {q^{n/2} - q^{-n/2}\over q^{1/2} - q^{-1/2}}.
\end{equation}

\end{document}